# Tight-Binding Study of Boron Structures

Joseph W. McGrady[1], Dimitrios A. Papaconstantopoulos[1], Michael J. Mehl[2]

[1]SPACS, George Mason University, Fairfax, VA, USA

[2]Naval Research Laboratory, Washington DC, USA

**Abstract**

We have performed Linearized Augmented Plane Wave (LAPW) calculations for five crystal structures (alpha, dhcp, sc, fcc, bcc) of Boron which we then fitted to a non-orthogonal tight-binding model following the Naval Research Laboratory Tight-Binding (NRL-TB) method.  The predictions of the NRL-TB approach for complicated Boron structures such as R105 (or β-rhombohedral) and T190 are in agreement with recent first-principles calculations.  Fully utilizing the computational speed of the NRL-TB method we calculated the energetic differences of various structures including those containing vacancies using supercells with up to 5000 atoms.

1. Introduction

Boron is of interest in materials science because it has at least 16 allotropes, making it one of the most structurally complex elements known [1].  Determining the ground state structure of Boron has been difficult, providing another source of motivation for pursuing research of this element.  However, it appears that in recent years the uncertainty about the ground structure of Boron may have begun shrinking.

In 2008, Widom and Mihalkovič determined that the slightly disordered β-rhombohedral (R105) structure is actually Boron's true ground state, as opposed to the α-rhombohedral (R12) structure [2].  They performed an optimization of the occupancy configurations of the R105 structure, however all possible configurations were not included.  VASP was used with the Local Density Approximation (LDA), as well as the Generalized Gradient Approximation (GGA)-based Ultrasoft Pseudopotential (USPP), and HARD potentials.  R105 was found to be lower in energy than R12 using USPP and HARD, but not LDA.  Zero-point energy was not included in the calculations, although they state that zero-point energy would make R105 even more stable.  Setten et al. found that R105 Boron is above R12 in energy unless zero-point energy is included [3].   There are crystallographic refinements of R105 Boron without any vacancies [4], but they do not reach the true ground state.

In a 2009 paper, Ogitsu et al. also found that R105 Boron is lower than the R12 structure after performing a full optimization of its occupancy configurations using a 1280 atom supercell [5].  Due to the very large number of possible occupancy configurations of the structure, only a small number of first principles calculations using the LDA were performed for a set of representative occupancy configurations.   These R105 total energies were then used to create a set of fitting coefficients based on the Ising model which allowed the total energy to be predicted based on the occupancy configuration.  Further, they used the symmetry and physical irrelevance of many of the configurations to reduce the number of calculations needed in the optimization.  Monte Carlo annealing simulations were performed in order to determine the stable configurations. The most stable structures were then further optimized with respect to the other lattice parameters. Zero-point energy was also included in their total energy results through ab-initio calculation of the phonon density of states of R12 and R105 Boron.

Oganov et al. summarize these recent findings by stating that the controversy over the ground state of Boron has been resolved, with R105 emerging as the true ground state [6]. Our results using an accurate and computationally efficient tight-binding approach reach the same conclusion.

## 2. Tight-Binding Method

We have used the NRL-TB method [7-9] in our exploration of the structure of Boron. NRL-TB is advantageous because of its fast performance when compared with first-principles methods such as the LAPW method [10]. In general, the NRL-TB method diagonalizes a 9Nx9N matrix for the s, p, and d orbitals, where N is the number of atoms in the unit cell. For Boron, we have omitted the d orbitals, so we only diagonalize a 4Nx4N matrix. NRL-TB's fast performance (it is about 1000 times faster than LAPW) makes the TB approach far more practical when dealing with structures with many atoms in the unit cell, as in the present work.

The NRL-TB method is based on a non-orthogonal version of the Slater-Koster two-center formalism [11] and uses a set of parameters fitted to first-principles total energy and energy band results in order to predict the total energies of structures which were not fitted. These parameters can be broken into the on-site parameters, the Hamiltonian parameters, and the overlap parameters. The parameter set is created by performing a non-linear least-squares fit to the first-principles data and determining a set of coefficients of two polynomials, listed below.

The on-site parameters depend on the orbital angular momentum and density of neighboring atoms, parametrized by the formula

$$h_{i\alpha} = a_{\bar{\iota}\alpha} + b_{\bar{\iota}\alpha}\rho_i^{2/3} + c_{\bar{\iota}\alpha}\rho_i^{4/3} + d_{\bar{\iota}\alpha}\rho_i^2 \qquad (1)$$

where $\rho_i$ is the atom density seen from atom i, and is given by the equation

$$\rho_i = \sum_{j \neq i} \exp[-\lambda^2 R_{ij}] F_c(R_{ij}) \qquad (2)$$

where $F_c(R_{ij})$ is a smooth cutoff function, and $R_{ij}$ is the distance between atoms i and j.

The Slater-Koster matrix elements of the Hamiltonian and overlap parameters are found from the equation

$$P_\gamma(R) = \left(e_\gamma + f_\gamma R + \overline{f}_\gamma R^2\right) \exp(-g_\gamma^2 R) F_c(R) \qquad (3)$$

where γ is the type of interaction (the interactions are ssσ, spσ, ppσ, ppπ in the case of Boron), R is the distance between atoms, and $F_c(R)$ is the same cutoff function as the one in equation (2). The coefficients $e_\gamma, f_\gamma, \overline{f}_\gamma,$ and $g_\gamma$ are different for the Hamiltonian and overlap matrices, but both have the functional form given in (3).

A shift in the first-principles (in our case LAPW) eigenvalues is performed that simplifies the total energy formula from Density Functional Theory (DFT) by making the sum of the eigenvalues equal to the total LAPW energy.

The DFT total energy expression is

$$E[n(r)] = \sum_i \epsilon_i + G[n(r)] \quad (4)$$

in which $\sum_i \epsilon_i$ is the sum of the eigenvalues over all k-points in the Brillouin zone and $G[n(r)]$ comprises the other terms for the DFT total energy.

New eigenvalues are created using the equation

$$\epsilon'_i = \epsilon_i + V_0 \quad (5)$$

in which the shift $V_0$ is given by the formula

$$V_0 = G[n(r)]/N_e \quad (6)$$

where $N_e$ is the total number of valence electrons and $n(r)$ is the electronic density. After performing the shift, the total LAPW energy is equal to the sum of the eigenvalues

$$E[n(r)] = \sum_i \epsilon'_i \quad (7)$$

The parameters are determined by a non-linear least-squares fit using a Levenberg-Marquardt algorithm [12, 13] which minimizes the mean-square error

$$M = \sum_i^j w_E(i)|E_{LAPW}(i) - E_{TB}(i)|^2 + \sum_{i,k,n} w_B(i,k,n)|\varepsilon_{LAPW}(i,k,n) - \varepsilon_{TB}(i,k,n)|^2 \quad (8)$$

where $E_{LAPW}(i)$ and $E_{TB}(i)$ are the total energies of the LAPW and Tight-Binding calculations for the i-th structure and $\varepsilon_{LAPW}(i,k,n)$ and $\varepsilon_{TB}(i,k,n)$ are the LAPW and TB eigenvalues, respectively, of the n-th band of the k-th k-point of the i-th structure. The weights $w_E(i)$ and $w_B(i,k,n)$ are chosen so that we can emphasize the relevant parts of the calculation. Typically $w_B$ is of order unity for bands near the Fermi energy, and $w_E$ is between 500 and 1000. The sums are over all structures i, over all k-points k for each structure, and over all valence/conduction bands n that are occupied or within about 1 Ry of the Fermi energy. The RMS error of the total energy fit of Equation (8) used to construct the parameter set given in Table A2.1 in Appendix 2 is 2mRy.

## 3. Total Energy Results

LAPW calculations for the sc, bcc, fcc, dhcp, and α-rhombohedral (R12) Boron structures were performed and used in a non-orthogonal NRL-TB method fitting to generate a set of 41 Tight-Binding coefficients. The coefficients were only calculated for the s and p orbitals. The d-states were omitted since Boron has only three s and p valence electrons.

The set of parameters listed in Table A2.1 was used to predict the total energies of 15 additional structures which were not fitted, many of which would be difficult to calculate with a first-principles method considering their size and complexity. A k-point mesh was generated for these Tight-Binding calculations, which was on an 8x8x8 grid.

The method of determining the total energy of a structure was to perform full optimizations for the structures. For structures with only one independently varying lattice parameter, this only required performing a volume optimization by calculating the total energy of the structure for a set of different volumes and obtaining the minimum energy. For structures with more than one independently varying lattice parameter, the entire process of volume optimization was repeated for multiple values of each additional independent parameter, and the minimum from all the configurations was taken as the true minimum total energy for the structure. The ground state predicted by these calculations is the hexagonal form of the R105 structure with one vacancy. Table 1 lists the minimum energy results for all of the total energy calculations performed, and the per-atom total energy vs. volume curves are shown in Figure 1(a).

**Table 1**. The per-atom volumes and total energies found for the structures explored are listed in columns 3 and 4, respectively. The table is ordered from lowest to highest total energy. In the first column, additional unit cell information such as angle and c/a is provided. The structures to which the parameters were fitted are marked as fitted in the first column.

| Structure | Space Group | Vol (Bohr^3/atom) | Total Energy (mRy/atom) |
|---|---|---|---|
| R105hex vacancy (c/a=2.18) | $P\bar{3}m1\text{-}D_{3d}^3$ | 51.392 | -104.72 |
| R105 (angle=64.74°) | $R\bar{3}m\text{-}D_{3d}^5$ | 51.078 | -104.63 |
| T190 (site configuration 588, c/a=1.4) | $P4_2/nnm\text{-}D_{4h}^{12}$ | 49.699 | -100.93 |
| γ-28 (b/a=1.11, c/a=1.39) | $Pnnm\text{-}D_{2h}^{12}$ | 46.955 | -100.61 |
| R12 (angle=58.5°) **(fitted)** | $R\bar{3}m\text{-}D_{3d}^5$ | 48.507 | -100.21 |
| T50 (c/a=0.57) | $P4_2/nnm\text{-}D_{4h}^{12}$ | 51.14 | -87.07 |
| aGa (c/a=1.1, angle=120.32°) | $Cmca\text{-}D_{2h}^{18}$ | 44 | -75.09 |
| betaSn (c/a=2.31) | $I4_1/amd\text{-}D_{4h}^{19}$ | 44.5 | -59.73 |
| c19 (angle=55.6°) | $R\bar{3}m\text{-}D_{3d}^5$ | 43 | -55.44 |
| DHCP (c/a=1.21) **(fitted)** | $P6_3/mmc\text{-}D_{6h}^4$ | 42.25 | -46.84 |
| A9 (c/a=1.31) | $P6_3/mmc\text{-}D_{6h}^4$ | 47.25 | -25.78 |
| C32 (c/a=1.02) | $P6/mmm\text{-}D_{6h}^1$ | 42.67 | -17.87 |
| shex (c/a=1.03) | $P6/mmm\text{-}D_{6h}^1$ | 42 | -15.19 |
| sc **(fitted)** | $Pm\bar{3}m\text{-}O_h^1$ | 44.739 | -11.51 |
| diam | $Fd\bar{3}m\text{-}O_h^7$ | 57.07 | 0.97 |
| aHg (angle=53.9°) | $R\bar{3}m\text{-}D_{3d}^5$ | 39 | 2.69 |
| A6 (c/a=1.29) | $I4/mmm\text{-}D_{4h}^{17}$ | 42 | 2.98 |
| fcc **(fitted)** | $Fm\bar{3}m\text{-}O_h^5$ | 39.366 | 4.06 |
| hcp (c/a=2.03) | $P6_3/mmc\text{-}D_{6h}^4$ | 39.5 | 8.02 |
| bMn | $P(4_1)32\text{-}O^7$ | 39.573 | 11.64 |
| bcc **(fitted)** | $Im\bar{3}m\text{-}O_h^9$ | 41.156 | 31.28 |

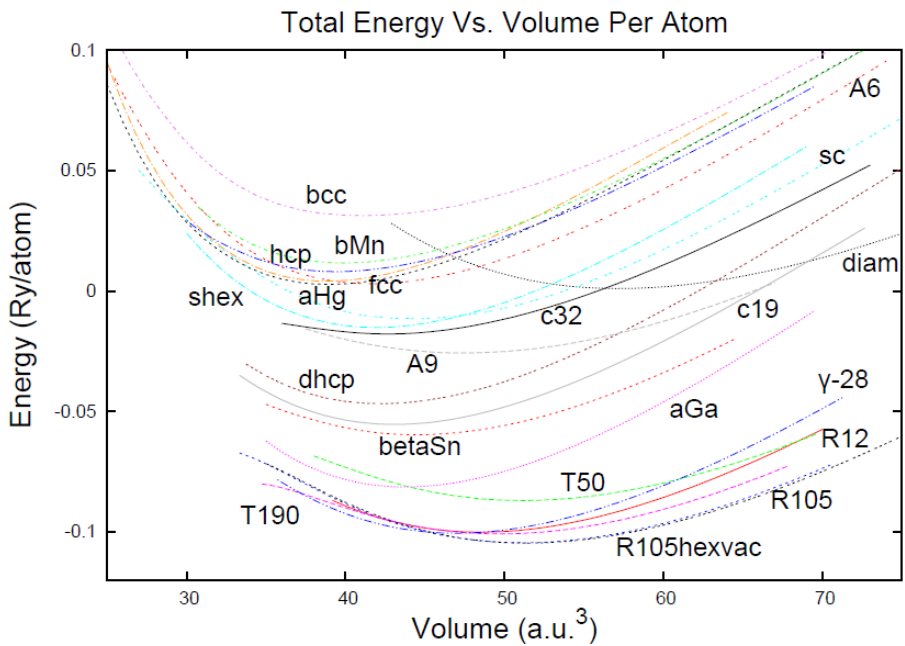

**Figure 1 (a).** The per-atom total energy is plotted as a function of per-atom unit cell volume for all of the structures listed in Table 1 with their minimum c/a and angle values. The T190 curve is for the minimum energy occupancy configuration found for the structure, and R105hexvac is the minimum vacancy form of the hexagonal R105 structure.

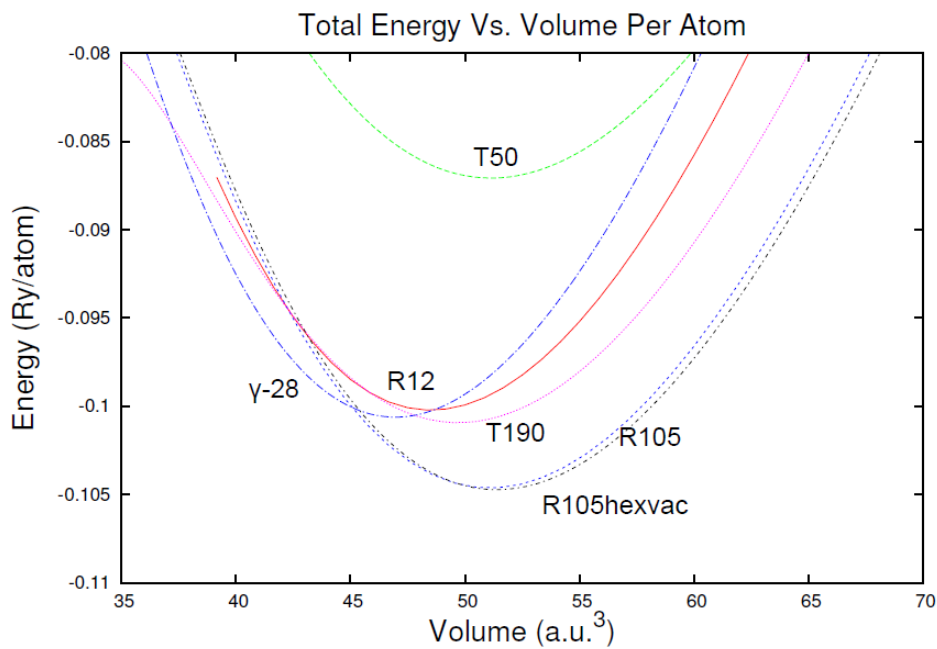

**Figure 1 (b).** The ground state region at the bottom of Figure 1(a) is shown in greater detail in order to emphasize the relationships among the most physically relevant structures of Boron. The energies of all structures shown except for R12 are predictions of the NRL-TB scheme.

The main Boron structures of interest are those which are lowest in energy: R12, R105, R105 hexagonal form, α-Tetragonal (T50), β-Tetragonal (T190), and γ-28. The five fitted structures are included in Table 1 and Figure 1(a) for comparison. 10 additional structures were predicted in order to demonstrate the capability of the NRL-TB method and the parameter set to predict higher energy structures that are far above the ground state region shown in Figure 1(b).

Boron R12 is rhombohedral with space group $R\bar{3}m$-$D_{3d}^5$ (No. 166) and 12 atoms in the unit cell. R12 was fitted when creating the parameters. NRL-TB gives a unit cell volume of 582 cubic Bohr and an angle of 58.5°. The experimental angle is 58.04°, with a unit cell volume of 590 cubic Bohr [14].

The hexagonal form of Boron R105 has space group $P\bar{3}m1$-$D_{3d}^3$ (No. 164) and 315 atoms in the unit cell. Details of the structure are presented in Appendix 1. Because of the computational speed of our method we performed all 315 calculations and then confirmed that the total energy results group according to the four symmetries of Appendix 1. R105hex becomes the NRL-TB predicted minimum energy structure with one vacancy at position (0.0, 0.0, 0.5) in lattice coordinates. In order to optimize the structure, we performed full optimizations with respect to volume and c/a for one vacancy at each of the 315 atom positions. Figure 2 displays the total energies of each of these vacancy calculations. Our predicted minimum c/a is 2.18, with a unit cell volume of 16137 cubic Bohr.

The rhombohedral form of Boron R105 has space group $R\bar{3}m$-$D_{3d}^5$ (No. 166) and 105 atoms in the unit cell. NRL-TB predicts an equilibrium Boron R105 volume of 5363 cubic Bohr, and the predicted angle Gamma is 64.74°. The experimental R105 angle is 65.32°, with a unit cell volume of 5539 cubic Bohr [1].

The Boron γ-28 structure is orthorhombic with space group $Pnnm$-$D_{2h}^{12}$ (No. 58) and 28 atoms in the unit cell. NRL-TB predicts that the lattice parameter ratios are b/a=1.11 and c/a=1.39, with a unit cell volume of 1315 cubic Bohr. The lattice constant ratios compare well with those previously predicted by Oganov et al., which are b/a=1.11, c/a=1.38 [15].

Boron T50 is a simple tetragonal lattice with space group $P4_2/nnm$-$D_{4h}^{12}$ (No. 134) and 50 atoms in the unit cell. Our NRL-TB prediction is c/a = 0.57, with a unit cell volume of 2557 cubic Bohr. The experimental c/a for this structure is 0.58, with a unit cell volume of 2590 cubic Bohr [1].

Boron T190 is a simple tetragonal lattice with space group $P4_2/nnm$-$D_{4h}^{12}$ (No. 134) with 190 atoms in the unit cell. Boron T190 was one of the most challenging structures to optimize, due to the fact that 12 of the 196 sites for this structure are partially occupied. In order to fully optimize the structure, it was assumed that 6 of the structure's 190 atoms would fill 6 of the 12 partial occupancy sites, leaving the other 6 empty. Volume optimizations with c/a held constant at 1.41 were then performed for all 924 possible ways of arranging 6 atoms in the 12 partial occupancy sites. The total energies resulting from each of these volume optimizations are displayed in Figure 3 below. The minimum energy configuration (number 588 on Figure 3) was further optimized by performing volume optimization for different c/a values. The minimum energy volume was 9443 cubic Bohr, with c/a = 1.4. The experimental volume given by Donohue is 9782 cubic Bohr, with c/a=1.397 [1]. Obviously, these 924-combination calculations became feasible because of the efficiency of the NRL-TB method and would be impractical to perform with standard DFT methods.

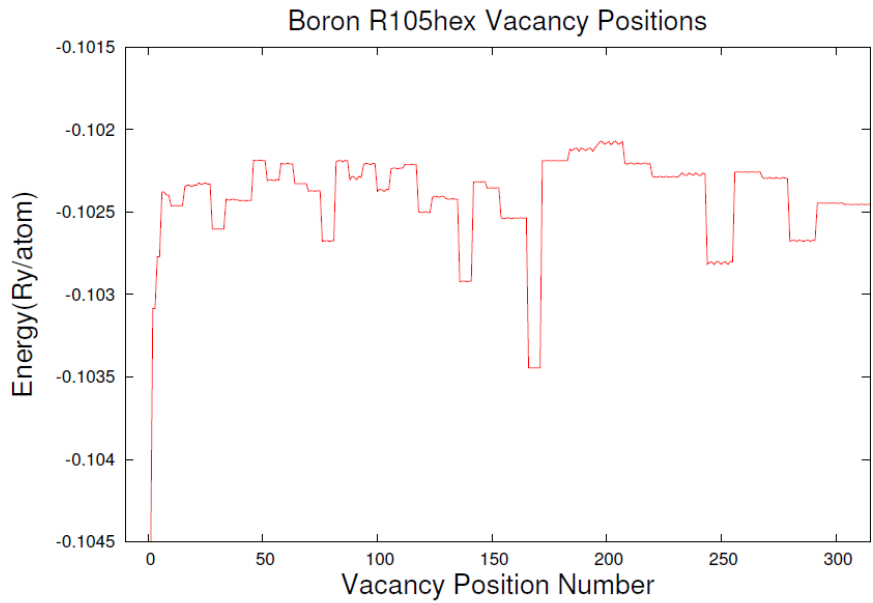

**Figure 2**.  The Boron R105 hexagonal per-atom total energy is plotted as a function of the number of the vacancy position.  The vacancy position number is an arbitrary index from 1 to 315 representing the atom position from which the atom is removed to create the vacancy.

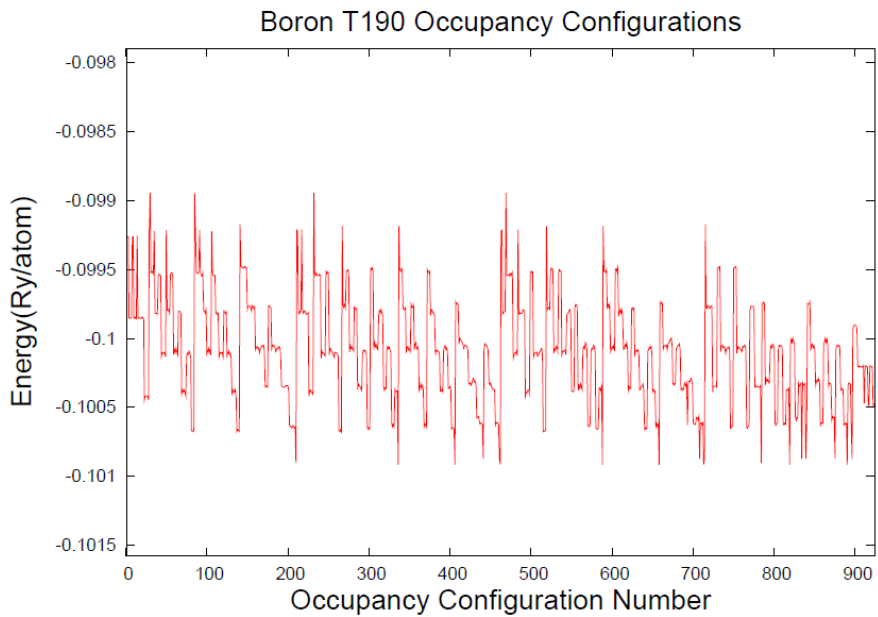

**Figure 3**.  The Boron T190 per-atom total energy is plotted as a function of the number of the occupancy configuration.  The occupancy configuration number is an arbitrary index from 1 to 924 representing the specific arrangement of 6 atoms in the 12 partially occupied sites.

## 4. Zero Point Energy

The zero-point energy was estimated for the low energy structures R12, R105, R105hex, T190, and γ-28. We used an approximation based on the Debye temperature [16-18] in which the zero-point energy is given by the equation

$$E_D = \frac{9}{8} k_B \Theta_D \tag{9}$$

where $k_B$ is the Boltzmann constant and $\theta_D$ is the Debye temperature. The formula used for the Debye temperature [18] is

$$\Theta_D = 67.48 \sqrt{\frac{rB}{M}} \tag{10}$$

where $r$ is the Wigner-Seitz radius in bohr, $B$ is the bulk modulus in kbar, and M is the mass in atomic mass units. The bulk modulus was determined using the equation

$$B = V \frac{d^2 E}{dV^2} \tag{11}$$

in which $E$ and $V$ are the total energies and volumes determined from a Birch-fit of our TB results.

Zero-point energy was calculated over a range of volumes and then added together with our original energies in order to determine new equilibrium volumes and energies for each structure. The ordering of the structures was not altered by the inclusion of zero-point energy, as can be seen by comparing the new equilibrium values displayed in Table 2 with the original total energies listed in Table 1. It's also interesting to note that the energy differences between the R105 hexagonal, R105, and the R12 structures are not changed substantially because they all have similar zero-point energy values. The equilibrium volumes for all the structures increased when zero-point energy was added, and Table 2 shows the percentage increase relative to the original volumes given in Table 1. The total energy vs. volume curves which incorporate zero-point energy are shown in Figure 4.

Equation (10) also allowed the Debye temperature to be determined in good agreement with experiment. The experimental Debye temperature for the R105 Boron structure is 1300K [19], which is consistent with the calculated values displayed in Table 2.

**Table 2**. The minimum total energies and volumes with zero-point energy included are shown along with the percentage by which the equilibrium volume increases after zero-point energy is added. The structures are ordered from lowest to highest total energy, showing that the same ordering as Table 1 is maintained when zero-point effects are taken into account. The structures have the same equilibrium lattice parameter values and occupancy configurations given in Table 1.

| Structure | Volume (bohr^3/atom) | Volume Increase | Total Energy with $E_D$ (mRy/atom) | $E_D$ (mRy/atom) | $\Theta_D$ (K) |
|---|---|---|---|---|---|
| R105hex vacancy | 51.730 | 0.7% | -95.4879 | 9.2237 | 1294.4933 |
| R105 | 51.521 | 0.9% | -95.3786 | 9.2251 | 1294.6881 |
| T190 | 50.556 | 1.7% | -92.1388 | 8.7073 | 1222.0257 |
| γ-28 | 47.431 | 1.0% | -90.7380 | 9.8403 | 1381.0260 |
| R12 | 49.006 | 1.0% | -90.6545 | 9.5273 | 1337.1078 |

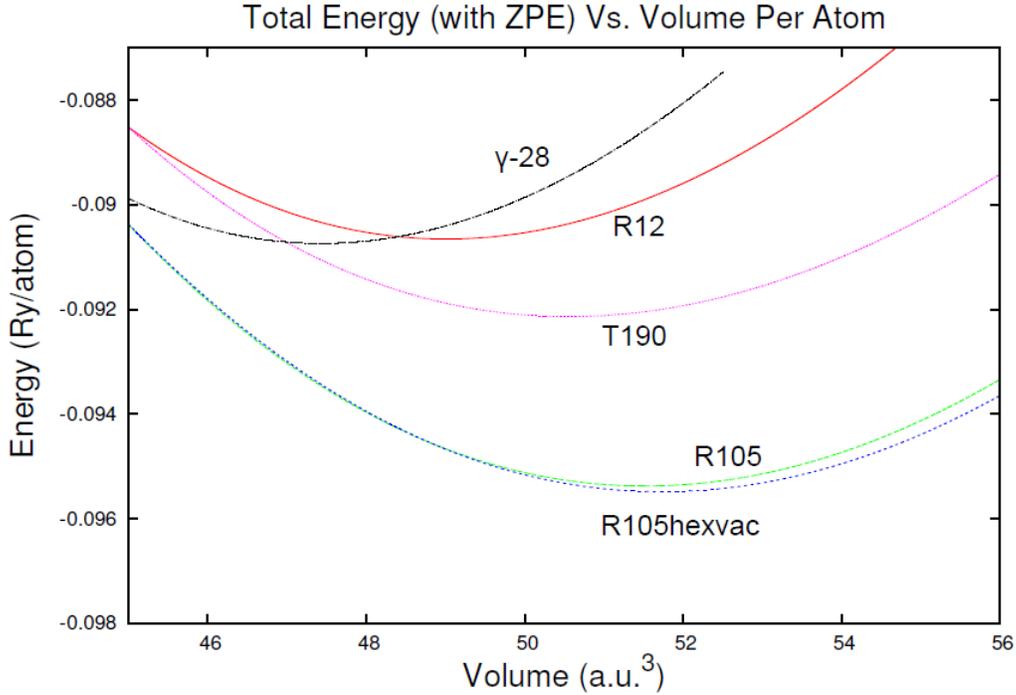

**Figure 4**.  The total energies of the lowest energy structures with zero-point energy added along the full curve.  The equilibrium volumes with zero-point energy are slightly larger, but the ordering of the structures doesn't change, as can be seen from comparison of this Figure with Figure 1 (b).

**5. Vacancy Formation Energies**

The set of Tight-Binding parameters used to predict total energies was not able to find relaxed vacancy formation energies due to the small nearest neighbor distances encountered in the conjugate gradient approach we used which led to an unphysical overlap matrix.  As a result, a new set of parameters was constructed from a fit to only the R12 Boron structure at multiple volumes in order to calculate relaxed vacancy formation energies.  These parameters are provided in Table A2.2.  These parameters also demonstrated transferability to the T190 and γ-28 structures, facilitating the calculation of relaxed vacancy formation energies for them as well.  Table 3 contains our results.

The formula used to calculate the vacancy formation energy is

$$E_{vac} = E_{vacancy} - \frac{(N-1)E_{bulk}}{N} \tag{12}$$

where $E_{bulk}$ and $E_{vacancy}$ are the total energies of the bulk and vacancy supercells, respectively, and N is the number of atoms in the supercell.

We have also calculated unrelaxed vacancy formation energies for each supercell for comparison.  These unrelaxed energies were calculated by taking $E_{vacancy}$ in (12) as the total energy of the vacancy supercell before any relaxation steps were performed.  As shown in Table 3, the unrelaxed energies are higher than the relaxed values.

For all vacancy formation energy calculations performed, only the Gamma k-point (0,0,0) was used, while in the total energy calculations shown in Table 1 a much larger mesh of 8x8x8 k-points was applied. This difference results in slight disagreement between the total energy given by the Tight-Binding Molecular Dynamics [20] software and the static Tight-Binding program used to predict the total energies. However, this difference diminishes as the size of the supercell increases, because the effect of a large supercell is similar to having more k-points with a smaller cell. The vacancy formation energies converge as the size of the supercell is increased, so multiple supercell sizes have been included in Table 3 to demonstrate this effect. We have not been able to find experimental results for the vacancy formation energies to compare with these calculations.

**Table 3**. This table contains relaxed and unrelaxed vacancy formation energies for multiple supercell sizes, as well as the nearest neighbor distance for each case. These results were calculated with the Tight-Binding Molecular Dynamics code using the parameter set shown in Table A2.2. The supercell size is given in terms of multiples of the unit cell sides, as well as the number of atoms. The equilibrium lattice parameters given in Table 1 were used for the structures.

| Structure | Supercell Used | Unrelaxed | | Relaxed | |
|---|---|---|---|---|---|
| | | Vac Energy (eV) | NN Dist (Bohr) | Vac Energy (eV) | NN Dist (Bohr) |
| R12 | 222 (96 atoms) | 2.53454 | 3.23140 | 2.01189 | 3.21848 |
| | 333 (324 atoms) | 2.59944 | 3.22538 | 2.01539 | 3.21977 |
| | 444 (768 atoms) | 2.74900 | 3.22703 | 2.18493 | 3.22191 |
| | 555 (1500 atoms) | 2.73198 | 3.22695 | 2.16766 | 3.22237 |
| | 666 (2592 atoms) | 2.60770 | 3.22686 | 2.04397 | 3.22235 |
| T190 | 111 (190 atoms) | 6.78671 | 3.09624 | 5.01396 | 3.12438 |
| | 222 (1520 atoms) | 7.37847 | 3.10849 | 5.04576 | 3.05714 |
| | 322 (2280 atoms) | 7.38875 | 3.10836 | 5.02916 | 3.05798 |
| | 332 (3420 atoms) | 7.40161 | 3.10836 | 5.03242 | 3.06050 |
| | 333 (5130 atoms) | 7.41752 | 3.10835 | 5.03857 | 3.06327 |
| γ-28 | 222 (224 atoms) | 3.57302 | 3.22241 | 2.20730 | 3.17587 |
| | 333 (756 atoms) | 4.01112 | 3.22763 | 2.50295 | 3.17203 |
| | 444 (1792 atoms) | 3.87530 | 3.22664 | 2.30962 | 3.16934 |
| | 555 (3500 atoms) | 3.98207 | 3.22691 | 2.42128 | 3.16844 |

## 6. Phonon Frequencies

The R12 phonon frequencies were calculated using the parameter set used for the vacancy formation energies, which is given in Table A2.2. The calculations were performed using the frozen phonon approximation as implemented in the ISOTROPY package [21]. We determine the energies for a set of structures identified by the package, which then determines the harmonic phonon frequencies and identifies phonon symmetries. We list in Table 4 the phonon frequencies at the Γ point for R12 Boron at equilibrium volume, and find good agreement with first-principles results we generated independently using VASP. Our parameter set did not predict the volume dependence of the phonon frequencies.

**Table 4.** Phonon frequencies for R12 Boron for several modes at the Γ point. The $\Gamma_3^+$ and $\Gamma_3^-$ modes are doubly degenerate. The TB calculations were done with the ISOTROPY package with the parameter set

fitted only to R12. VASP results for each case are provided to demonstrate the quality of the TB predictions. The units are inverse cm.

| $\Gamma_1^+$ | | $\Gamma_2^+$ | | $\Gamma_3^+$ | | $\Gamma_1^-$ | | $\Gamma_2^-$ | | $\Gamma_3^-$ | |
|---|---|---|---|---|---|---|---|---|---|---|---|
| VASP | TB | VASP | TB | VASP | TB | VASP | TB | VASP | TB | VASP | TB |
| 709 | 810 | 500 | 237 | 520 | 381 | 504 | 580 | 0 | 0 | 0 | 0 |
| 825 | 844 | 743 | 780 | 612 | 629 | 828 | 997 | 826 | 855 | 578 | 611 |
| 947 | 1055 | | | 740 | 726 | | | 845 | 1000 | 614 | 639 |
| 1216 | 1154 | | | 801 | 856 | | | 979 | 1049 | 720 | 703 |
| | | | | 901 | 973 | | | | | 828 | 1012 |
| | | | | 1159 | 1119 | | | | | 832 | 1046 |

## 7. Elastic Constants

We calculated the elastic constants for R12 Boron using the same parameter set used for the phonon frequencies and vacancy formation energies. The calculations were performed using the finite strain method [22], choosing appropriate strains to determine the six independent elastic constants for the rhombohedral structure, and allowing the atoms to relax once the primitive cell is strained appropriately. As shown in Table 5, our TB predicted constants are consistent with the first-principles VASP results we calculated for comparison.

**Table 5.** The elastic constants for the R12 Boron structure calculated with both VASP and TB. In the rhombohedral system, $C_{24} = -C_{14} = C_{56}$, and $C_{66} = (C_{11} - C_{12})/2$. The values are given in units of GPa.

| Elastic Constant | VASP | TB |
|---|---|---|
| $C_{11}$ | 488 | 509 |
| $C_{12}$ | 117 | 155 |
| $C_{33}$ | 633 | 650 |
| $C_{13}$ | 51 | 68 |
| $C_{44}$ | 213 | 154 |
| $C_{14}$ | -26 | -21 |

## 8. Density of States

An additional set of parameters was fitted to LAPW bands for the experimental Boron R12 structure [23] in order to perform Tight-Binding electronic density of states (DOS) calculations. This set is shown in Table A2.3. The calculation was done for a unit cell volume of 555 cubic Bohr, and angle 58.16°. The Tight-Binding method was able to accurately recreate the LAPW DOS for the same experimental lattice parameters. It demonstrates a band gap and the same positions of the DOS peaks, as can be seen in Figure 5. It should also be mentioned here that Ref. 23 discusses superconductivity under pressure for a metallic fcc phase of Boron which was also fitted in the present work.

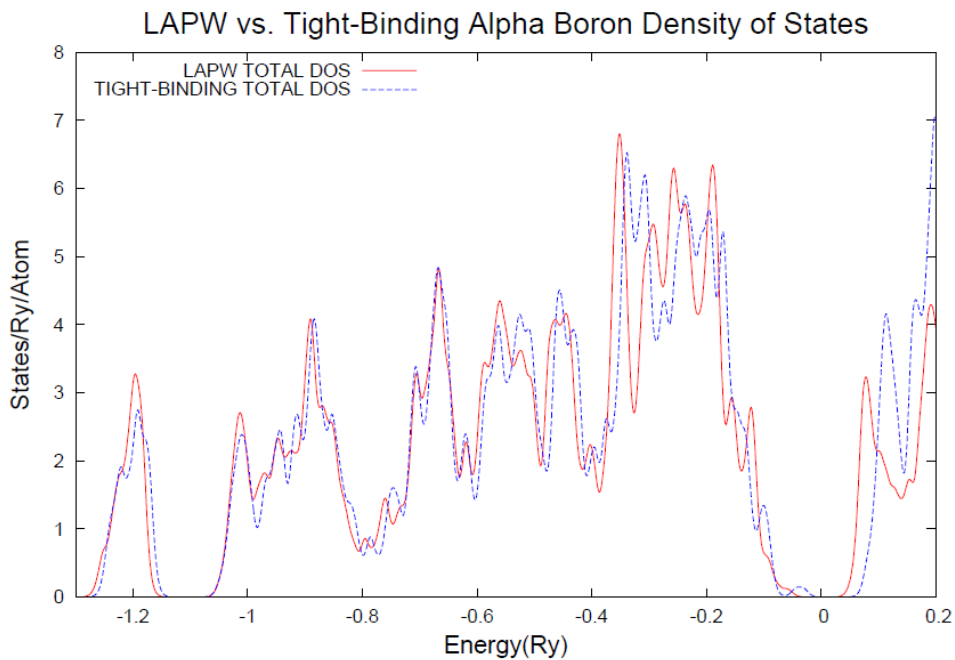

**Figure 5**. The number of states per Rydberg per atom is plotted for both the LAPW and NRL-TB methods. The NRL-TB energy was shifted in order to make the Fermi level consistent with the LAPW results so that the two plots could be displayed together.

## 9. Conclusions

The NRL-TB method predicts the R105 hexagonal structure, with one vacancy, as the ground state of Boron based on a fit to original first-principles LAPW results, which further strengthens the consensus that R105 is the true ground state of Boron. An optimization of the vacancy position of R105 was performed, and the lattice parameters in each case were fully optimized. NRL-TB also facilitated extensive calculations for the T190 structure covering all possible occupancy configurations and giving the correct result that T190 has higher total energy than R105. When an approximation for the zero-point energy was included, the predicted ordering of the lowest energy structures remained unchanged. The vacancy formation energy, phonon frequency and elastic constant calculations required a second parameter set fitted to the energy bands and total energies of the R12 structure. This parameter set also found vacancy formation energies for the T190 and γ-28 structures, which were not fitted. The TB R12 phonon frequencies at equilibrium volume compare well with first principles results from VASP, however the volume dependence of the phonon frequencies could not be predicted by our parameters. We developed a third set of parameters fitted solely to R12 bands to calculate the Boron R12 DOS, successfully reproducing the DOS peaks and the gap.

**Acknowledgment:** Research funded in part by DoE grant DE-FG02-07ER-46425 and by ONR grant N00014-09-1-1025

## References

[1] J. Donohue, "*The Structures of the Elements.*" Krieger Pub Co (1982).

[22] M. J. Mehl, J. E. Osburn, D. A. Papaconstantopoulos, and B. M. Klein, Phys. Rev. B 41, 10311 (1990); erattum Phys. Rev. B. 42, 5362 (1990); M. J. Mehl, B. M. Klein, D. A. Papaconstantopoulos. "Chapter 9: First-Principles Calculation of Elastic Properties." *Intermetallic Compounds*: Vol 1, *Principles*. Edited by J. H. Westbrook and R. L. Fleischer. John Wiley and Sons Ltd, 1994. 195-210. Print.

[23] D. A. Papaconstantopoulos, M. J. Mehl. 2002. *Phys. Rev*. B 65 172510.

**Appendix 1:** Vacancies in the β-Boron Structure

As noted in the introduction to the main paper, Widom and Mihalkovič [2] have shown that an ordered pattern of vacancies in the R105 (β-Boron) structure makes that structure a likely candidate for the T=0 ground state of crystalline Boron. While we will not study their exact structures, in this appendix we show that our tight-binding parameters do, indeed, produce low energy structures when atoms are removed from the fully-occupied R105 structure.

We begin with the fully occupied R105 structure of β-Boron reported by Geist [3]. This is a rhombohedral unit cell, space group $R\bar{3}m$-$D_{3d}^5$ (No. 166), Pearson Symbol hR105. The (1b), (2c), nine independent (6h) and four independent (12i) Wyckoff positions are occupied, giving fifteen independent classes of atoms.

We construct our vacancy cell as follows: we first convert the 105-atom rhombohedral unit cell to its equivalent 315-atom hexagonal representation. We then remove, in order, one atom from each of the Wyckoff positions listed above, and compute the energy of the cell without relaxation. Symmetry ensures that this covers all possible 314 atom structures that can be calculated by this method. We thus generate the following types of lattices:

1. Removing a (1b) atom from the 315 atom unit cell produces a crystal structure with space group $P\bar{3}m1$-$D_{3d}^3$ (No. 164), Pearson Symbol hP314, with atoms on the Wyckoff (2c) site, three different (2d) sites, twenty-seven different (6i) sites, and twelve different (12j) sites.
2. Removing a (2c) atom results in a cell with space group $P3m1$-$C_{3v}^1$ (No. 156), Pearson Symbol hP314, with atoms on two different Wyckoff (1a) sites, three different (1b) sites, three different (1c) sites, fifty-four different (3d) sites, and twenty-four different (6e) sites.
3. Removing a (6h) atom yields a cell with space group $Cm$-$C_s^3$ (No. 8), Pearson Symbol mC628, with sixty-two (2a) sites and one hundred twenty-six (4b) sites. Note that there are four possible atoms that can be removed here.
4. Removing a (12i) atom results in a unit cell with essentially no symmetry: the space group is $P1$-$C_1^1$ (No. 1), Pearson Symbol aP314, and all three hundred fourteen atoms are on independent (1a) sites. Since there are four different (12i) sites, there are four different possible structures.

As noted in the main text, the lowest energy structure of all these is the one described in item 1 above, which is slightly lower than the energy of the perfectly ordered hR105 structure.

**Appendix 2:** NRL Tight-Binding Parameter Sets

The 8 on site coefficients and the 32 Hamiltonian and overlap coefficients are listed in these tables. Each header section contains the functional form for the parameters in that category. Each of the equations has units of energy (Ry), so the parameters must have the appropriate units to ensure this.

**Table A2.1**. These parameters were used to calculate the total energies given in Table 1 of the main text, and they were fitted to LAPW total energies for the sc, bcc, fcc, dhcp, and R12 Boron structures. The value of lambda is 1.11382275935.

| On-site parameters $h_l = a_l + b_l \rho^{2/3} + c_l \rho^{4/3} + d_l \rho^2$ | | | | |
|---|---|---|---|---|
| $l$ | $a_l$ (Ry) | $b_l$ (Ry) | $c_l$ (Ry) | $d_l$ (Ry) |
| $s$ | 0.424836994512E-01 | 0.458208054422E-01 | -0.149821938058E+01 | 0.276689175071E+01 |
| $p$ | 0.548658121707E+00 | -0.391380983048E-01 | -0.105900874054E+01 | 0.177939103639E+01 |
| Hamiltonian parameters $H_{ll'u}(R) = (e_{ll'u} + f_{ll'u}R + \overline{f}_{ll'u}R^2) \exp(-g_{ll'u}^2 R) F_c(R)$ | | | | |
| $H_{ll'u}$ | $e_{ll'u}$ (Ry) | $f_{ll'u}$ (Ry/Bohr) | $\overline{f}_{ll'u}$ (Ry/Bohr$^2$) | $g_{ll'u}$ (Bohr$^{-1/2}$) |
| $H_{sss}$ | -0.207727375107E+02 | 0.104720879238E+02 | -0.203340035822E+01 | 0.110609250706E+01 |
| $H_{sps}$ | 0.486401985504E+00 | -0.201325444832E+01 | 0.890894174293E+00 | 0.100121654776E+01 |
| $H_{pps}$ | 0.811946083218E+01 | -0.687054115892E+01 | 0.219263663490E+01 | 0.105072214525E+01 |
| $H_{ppp}$ | 0.140024687567E+02 | 0.327327604185E+01 | -0.302945766423E+01 | 0.116512314866E+01 |
| Overlap parameters $S_{ll'u}(R) = (e'_{ll'u} + f'_{ll'u}R + \overline{f'}_{ll'u}R^2) \exp(-g'^2_{ll'u} R) F_c(R)$ | | | | |
| $S_{ll'u}$ | $e'_{ll'u}$ | $f'_{ll'u}$ (1/Bohr) | $\overline{f'}_{ll'u}$ (1/Bohr$^2$) | $g'_{ll'u}$ (Bohr$^{-1/2}$) |
| $S_{sss}$ | -0.513069625260E+01 | -0.109229602472E+01 | 0.158331362883E+01 | 0.102841788222E+01 |
| $S_{sps}$ | 0.257252096552E+02 | -0.380835149672E+01 | -0.309857944656E+01 | 0.117015437917E+01 |
| $S_{pps}$ | 0.647210209217E+02 | -0.138913540657E+02 | -0.421559000803E+01 | 0.123100905029E+01 |
| $S_{ppp}$ | 0.165945231405E+02 | -0.592184916271E+01 | 0.494764628228E+00 | 0.972511418727E+00 |

**Table A2.2**. This parameter set was fitted to total energies and bands of the R12 Boron structure and was used in the prediction of the vacancy formation energies, phonon frequencies and elastic constants given in Tables 3, 4 and 5 of the main text. The value of lambda for this parameter set is 1.11379367927.

| On-site parameters $h_l = a_l + b_l \rho^{2/3} + c_l \rho^{4/3} + d_l \rho^2$ | | | | |
|---|---|---|---|---|
| $l$ | $a_l$ (Ry) | $b_l$ (Ry) | $c_l$ (Ry) | $d_l$ (Ry) |
| $s$ | 0.423171678675E-01 | 0.578783864122E-01 | -0.141353037846E+01 | 0.317762742556E+01 |
| $p$ | 0.547912154116E+00 | -0.398987110109E-01 | -0.105811401503E+01 | 0.177616196305E+01 |
| Hamiltonian parameters $H_{ll'u}(R) = (e_{ll'u} + f_{ll'u}R + \overline{f}_{ll'u}R^2) \exp(-g_{ll'u}^2 R) F_c(R)$ | | | | |
| $H_{ll'u}$ | $e_{ll'u}$ (Ry) | $f_{ll'u}$ (Ry/Bohr) | $\overline{f}_{ll'u}$ (Ry/Bohr$^2$) | $g_{ll'u}$ (Bohr$^{-1/2}$) |
| $H_{sss}$ | -0.251636277521E+02 | 0.116619408301E+02 | -0.178566136933E+01 | 0.111514619799E+01 |
| $H_{sps}$ | 0.495145075264E+00 | -0.201043096340E+01 | 0.891277389103E+00 | 0.100665817385E+01 |
| $H_{pps}$ | 0.812557167785E+01 | -0.686730287481E+01 | 0.219413838327E+01 | 0.105027892799E+01 |
| $H_{ppp}$ | 0.137060713311E+02 | 0.318571625108E+01 | -0.305511064124E+01 | 0.116183752583E+01 |
| Overlap parameters $S_{ll'u}(R) = \left(e'_{ll'u} + f'_{ll'u}R + \overline{f'}_{ll'u}R^2\right) \exp\left(-g'^2_{ll'u}R\right) F_c(R)$ | | | | |
| $S_{ll'u}$ | $e'_{ll'u}$ | $f'_{ll'u}$ (1/Bohr) | $\overline{f'}_{ll'u}$ (1/Bohr$^2$) | $g'_{ll'u}$ (Bohr$^{-1/2}$) |
| $S_{sss}$ | -0.285204015456E+01 | -0.567665493485E+00 | 0.165987874417E+01 | 0.105347613633E+01 |
| $S_{sps}$ | 0.254646235160E+02 | -0.378086606738E+01 | -0.308786814776E+01 | 0.117040580615E+01 |
| $S_{pps}$ | 0.103867472453E+03 | -0.637905089182E+01 | -0.173935185100E+01 | 0.149054074564E+01 |
| $S_{ppp}$ | 0.161535776273E+02 | -0.604361078734E+01 | 0.462384819057E+00 | 0.101065538838E+01 |

**Table A2.3**. This parameter set was fitted to only the bands of the R12 Boron structure and was used in the prediction of the R12 DOS shown in Figure 5 in the main text. The value of lambda for this parameter set is 1.05396527367.

| On-site parameters $h_l = a_l + b_l\rho^{2/3} + c_l\rho^{4/3} + d_l\rho^2$ | | | | |
|---|---|---|---|---|
| $l$ | $a_l$ (Ry) | $b_l$ (Ry) | $c_l$ (Ry) | $d_l$ (Ry) |
| s | -0.288944565362E+01 | -0.337654603710E+01 | 0.568689682985E+01 | 0.864452562914E+02 |
| p | -0.601795903871E+00 | -0.126582861414E+01 | 0.199204480099E+01 | 0.316825989417E+02 |
| Hamiltonian parameters $H_{ll'u}(R) = \left(e_{ll'u} + f_{ll'u}R + \overline{f}_{ll'u}R^2\right) \exp\left(-g^2_{ll'u}R\right) F_c(R)$ | | | | |
| $H_{ll'u}$ | $e_{ll'u}$ (Ry) | $f_{ll'u}$ (Ry/Bohr) | $\overline{f}_{ll'u}$ (Ry/Bohr$^2$) | $g_{ll'u}$ (Bohr$^{-1/2}$) |
| $H_{sss}$ | -0.630900136603E+01 | -0.218323900060E-01 | 0.882511491627E-01 | 0.985597810912E+00 |
| $H_{sps}$ | 0.206951221103E+01 | 0.150439870118E+01 | -0.160390161549E+00 | 0.100042653784E+01 |
| $H_{pps}$ | -0.279851321149E+02 | 0.401708483324E+01 | 0.273181142387E+01 | 0.114839211586E+01 |
| $H_{ppp}$ | 0.476925197262E+03 | 0.157034915119E+01 | -0.129623735837E+03 | 0.171986690302E+01 |
| Overlap parameters $S_{ll'u}(R) = \left(e'_{ll'u} + f'_{ll'u}R + \overline{f'}_{ll'u}R^2\right) \exp\left(-g'^2_{ll'u}R\right) F_c(R)$ | | | | |
| $S_{ll'u}$ | $e'_{ll'u}$ | $f'_{ll'u}$ (1/Bohr) | $\overline{f'}_{ll'u}$ (1/Bohr$^2$) | $g'_{ll'u}$ (Bohr$^{-1/2}$) |
| $S_{sss}$ | 0.257713343182E+01 | 0.602338767490E+00 | 0.390008071028E+00 | 0.114901014672E+01 |
| $S_{sps}$ | 0.279580102847E+01 | -0.264174873169E+01 | 0.341371874239E+00 | 0.936215374169E+00 |
| $S_{pps}$ | 0.113914681055E+04 | -0.908769292021E+02 | -0.943436347904E+02 | 0.153636956486E+01 |
| $S_{ppp}$ | -0.163620121706E+04 | 0.166510038428E+03 | 0.113746759651E+03 | 0.149279638374E+01 |